\documentclass[aps, superscriptaddress, showpacs, floatfix, twocolumn, longbibliography]{revtex4-1}

\usepackage{amsmath}
\usepackage{amssymb}
\usepackage{graphicx}
\usepackage{epstopdf}
\usepackage{color}
\usepackage{soul}
\usepackage{hyperref}

\hypersetup{
  colorlinks = true, 
  urlcolor   = blue, 
  linkcolor  = blue, 
  citecolor  = blue  
}

\begin{document}

\title{Correlation length in a generalized two-dimensional XY model}
\author{Duong Xuan Nui}
\affiliation{Advanced Institute for Science and Technology, Hanoi University of Science and Technology, Hanoi 10000, Vietnam}
\affiliation{Faculty of Mechanics and Civil Engineering, Vietnam National University of Forestry, Xuan Mai, Chuong My district, Hanoi 10000, Vietnam}
\author{Le Tuan}
\affiliation{School of Engineering Physics, Hanoi University of Science and Technology, Hanoi 10000, Vietnam}
\author{Nguyen Duc Trung Kien}
\affiliation{Advanced Institute for Science and Technology, Hanoi University of Science and Technology, Hanoi 10000, Vietnam}
\author{Pham Thanh Huy}
\affiliation{Advanced Institute for Science and Technology, Hanoi University of Science and Technology, Hanoi 10000, Vietnam}
\author{Hung T. Dang}
\affiliation{Advanced Institute for Science and Technology, Hanoi University of Science and Technology, Hanoi 10000, Vietnam}
\affiliation{Thanh Tay Institute for Advanced Study (TIAS), Thanh Tay University, Yen Nghia, Ha-Dong district, Hanoi 10000, Vietnam}
\affiliation{Phenikaa Research and Technology Institute (PRATI), A\&A Green Phoenix Group, 167 Hoang Ngan, Hanoi 10000, Vietnam}
\author{Dao Xuan Viet}
\affiliation{Advanced Institute for Science and Technology, Hanoi University of Science and Technology, Hanoi 10000, Vietnam}
\date{\today}

\begin{abstract}
The measurements of the magnetic and nematic correlation lengths in a generalization of the two-dimensional XY model on the square lattice are presented using classical Monte Carlo simulation. The full phase diagram is reexamined based on these correlation lengths, demonstrating their power in studying generalized XY models. The ratio between the correlation length and the lattice size has distinctive behaviors which can be used to distinguish different types of phase transition. More importantly, the magnetic correlation length give more insights into the tricritical region where the paramagnetic, nematic and quasi-long-range phases meet. It shows signatures for the intermediate region starting from the tricritical point, where the transition line is neither of the same physics as the Ising transition below nor the Berezinskii-Kosterlitz-Thouless transition far above the tricritical point.
\end{abstract}

\maketitle

\section{Introduction \label{sec:intro}}

In short-range interacting systems, two dimension is the lower critical dimension where continuous symmetry breaking is not able to occur, as stated by the Mermin-Wagner theorem \cite{Mermin1966}. However, it is the marginal dimension where one can observe instead topological changes. Particularly in the XY model, vortices and antivortices, which are topological excitations, lead to a phase transition from the disordered phase of free vortices at high temperatures, where the distance-dependent spin-spin correlation function decays exponentially, to a low-temperature phase of quasi-long-range order of pairs of bound vortices where that function decays algebraically \cite{Berezinskii1971,Kosterlitz1973}, namely the Berezinskii-Kosterlitz-Thouless (BKT) transition.

Several generalizations of the XY model have been proposed in order to search for novel phenomena that can be realized in physical experiments or realistic materials \cite{Korshunov1985,Lee1985,Domany1984,Romano2002,Taroni2008}. Since 1985, starting with the works of Korshunov \cite{Korshunov1985}, Lee and Grinstein \cite{Lee1985}, the generalized XY models which include nematic effects have gained much attention because of their possibility for investigating BKT phase transitions in liquid crystal \cite{Pang1992,Lee1985}, bosonic mixtures in ultracold atomic/molecular systems or in He$^3$ thin films \cite{Korshunov1985,Bonnes2012,Bhaseen2012}. In these models, together with the original magnetic interaction with spin angle periodicity of $2\pi$, there is an extra nematic interaction characterized by a positive integer $q$ such that its periodicity is $2\pi/q$. As a result, besides the conventional vortices and antivortices generated by the magnetic interaction, there are $1/q$-integer vortices which are the products of the nematic interaction and have a noninteger ($1/q$) winding number.

Depending on the relative strength between these two interactions, the above generalized XY model experiences different phases. When the magnetic interaction is dominant, it becomes the conventional XY model where there is only a BKT transition from the disordered paramagnetic phase to the quasi-long-range order. In contrast, if the nematic interaction is dominant, a nematic phase can be stabilized at low temperature where, similar to the quasi-long-range order, there are bound pairs of noninteger vortices. When both interactions contribute, the physics is different for different $q$'s. At $q < 4$, there are three possible phases in the phase diagrams that meet at a tricritical point: the disordered (paramagnetic) phase (P), the quasi-long-range ordered phase (F), and the nematic phase (N) \cite{Korshunov1985,Lee1985,Carpenter1989,Poderoso2011,Canova2014,Canova2016}. Away from the multicritical region, the phase transitions from the disordered to the nematic or the quasi-long-range phase belong to the BKT universality class \cite{Carpenter1989,Huebscher2013}, while the transition from the nematic to the quasi-long-range order belongs to the Ising university class for $q=2,4$ \cite{Poderoso2011,Canova2016} or to the Pott universality class for $q=3$ \cite{Poderoso2011,Canova2014}. For $q\ge 4$, Refs.~\onlinecite{Poderoso2011,Canova2016} show that there are two new phases which, together with the quasi-long-range order, can be differentiated by the angle distribution of the spins.

In this work, we focus on the case $q=2$ of this generalized two-dimensional XY model. Its phase diagram has been constructed since the early days of the model \cite{Korshunov1985,Lee1985,Carpenter1989} and most of the physics are now rather well understood. The focus at present is around the tricritical point where all the phase boundaries meet. The remaining issue is whether it is a true tricritical point where all the phase transition lines end at this point. Several recent works have shown that the Ising line for the transition from the nematic to the quasi-long-range order can extend beyond the tricritical point \cite{Shi2011,Huebscher2013,Serna2017}, so that there is a segment of the phase boundary where the transition directly from the quasi-long-range phase to the disordered phase belongs to the Ising universality class. Reference~\onlinecite{Serna2017} regarded it as a classical example of the deconfined quantum criticality \cite{Senthil2004}. In this paper, by employing large-scale Monte Carlo simulations, we study in detail the behaviors of the correlation length. This quantity has been investigated for various models (see Ref.~\onlinecite{Salas2000} and references therein). However, to our knowledge, it has not been studied rigorously for the generalized XY model of our interest. Therefore, in this study, we provide a detailed analysis of the nematic and magnetic correlation lengths, based on which the full phase diagram of the model is reproduced. More importantly, with these measurements, we give more insights into the physics of the tricritical region, in particularly the understanding of the Ising phase boundary originating from the nematic-quasi-long-range order transition, how the line ends and how the system changes from the Ising phase transition to the BKT transition when the relative strength of the interactions is varied.

The structure of this paper is as follows. Section~\ref{sec:model_methods} presents the methods used in this study. In Sec.~\ref{sec:phase_diagram} we discuss the behaviors of the correlation length ratios $\frac{\xi}{L}$ for different phases, then reconstruct the full phase diagram of the model. In Sec.~\ref{sec:ising_transition} we study the phase transition in the tricritical region based on the correlation length measurements. Section~\ref{sec:conclusions} concludes our study.

\section{Model and Methods \label{sec:model_methods}}

The classical Hamiltonian of the two-dimensional generalized XY model under consideration has two terms: the usual ferromagnetic term and the extra $q$-nematic term. The contribution of each term is controlled by a parameter $\Delta$ running from $0$ to $1$. The specific form of the Hamiltonian reads

\begin{equation}
 \label{eq:generalized_xy_ham}
 H = -\sum_{\langle ij\rangle} [\Delta \cos(\theta_i-\theta_j) + (1-\Delta) \cos(q\theta_i-q\theta_j)].
\end{equation}
The angle $\theta_i$ ($\theta_j$) is made of the spin direction located at the site $i$ ($j$) with the $x$ axis. The interactions are short-ranged, occuring between nearest neighbor sites, denoted by $\langle ij\rangle$ in Eq.~\eqref{eq:generalized_xy_ham}. The basic energy scale is the ferromagnetic coupling, which is set to $1$. We focus only on the case $q=2$ in this study.

We choose the square lattice  of size $L$ in each direction, thus there are $N=L^2$ sites in total. The periodic boundary condition is applied in both directions. The size $L$ for simulations is chosen from $16$ to $256$, measurements in the thermodynamic limit are obtained by extrapolating the data from simulations within this range of $L$.

Monte Carlo method is employed in our study with two types of updates: local single-spin-flip Metropolis algorithm and cluster-spin-flip algorithm following the Wolff algorithm \cite{Wolff1989}. Local and cluster updates are carried out once in every Monte Carlo step. For each case we perform five runs, each with a different random seed. For each run, there are $4\times10^6 \to 6\times10^6$ updates for equalibration and $6\times 10^6$ updates for measurements. We check the equality of the specific heat computed via the energy fluctuation and that calculated via the temperature difference of the energy to ensure that the system is in equilibrium.

For $q=2$, we carry out Monte Carlo simulations to measure the second-moment correlation length \cite{Salas2000,Zierenberg2017,Viet2009,Viet2009b}, of which we consider two types:  the magnetic one ($\xi_1$) and the nematic one ($\xi_2$). The general form reads
\begin{equation}
\label{eq:correlation_length}
 \xi_n = \dfrac{1}{2 \sin(k_\mathrm{m}/2)} \sqrt{\dfrac{ \langle m_n (\vec 0)^2 \rangle } {\langle m_n (\vec{k}_\mathrm{m})^2 \rangle } -1 }, \\
\end{equation}
where $\vec{k}_m = (2\pi/L, 0)$ and $\langle \cdots \rangle$ denotes the thermal average. The $k$-dependent magnetization is
\begin{equation}
m_n(\vec k)^2 = \sum_{\mu =x,y}\left|\dfrac {1}{N} \sum_{i=1}^N S^n_{i\mu} \exp(i\vec k \cdot \vec r_{i})\right|^2.
\end{equation}
The projections of the spin to the $x$ and $y$ axes are $S^n_{ix}=\cos(n\theta_i)$ and $S^n_{iy} = \sin(n\theta_i)$. In the thermodynamic limit, the correlation lengths diverge when the system goes into the (quasi-long-range) ordered phase, thus the second-moment correlation length [Eq.~\eqref{eq:correlation_length}] is related to the true correlation length (derived from the correlation function) only for $T>T_c$ \cite{Ballesteros2000}. On the other hand, for finite-size systems, $\xi_n$ scales with the lattice size $L$ at criticality, thus the ratios $\frac{\xi_n}{L}$ at $T=T_c$ remains finite and is claimed to be universal \cite{Salas2000}. They are the main objects of our investigation.

We also measure the specific heat (presented in Appendix~\ref{app:specific_heat}) which is calculated based on the variance of the total energy to supplement the results provided by the correlation lengths.

\section{Phase diagram \label{sec:phase_diagram}}

\begin{figure}[t]
\centering
\includegraphics[width=\columnwidth]{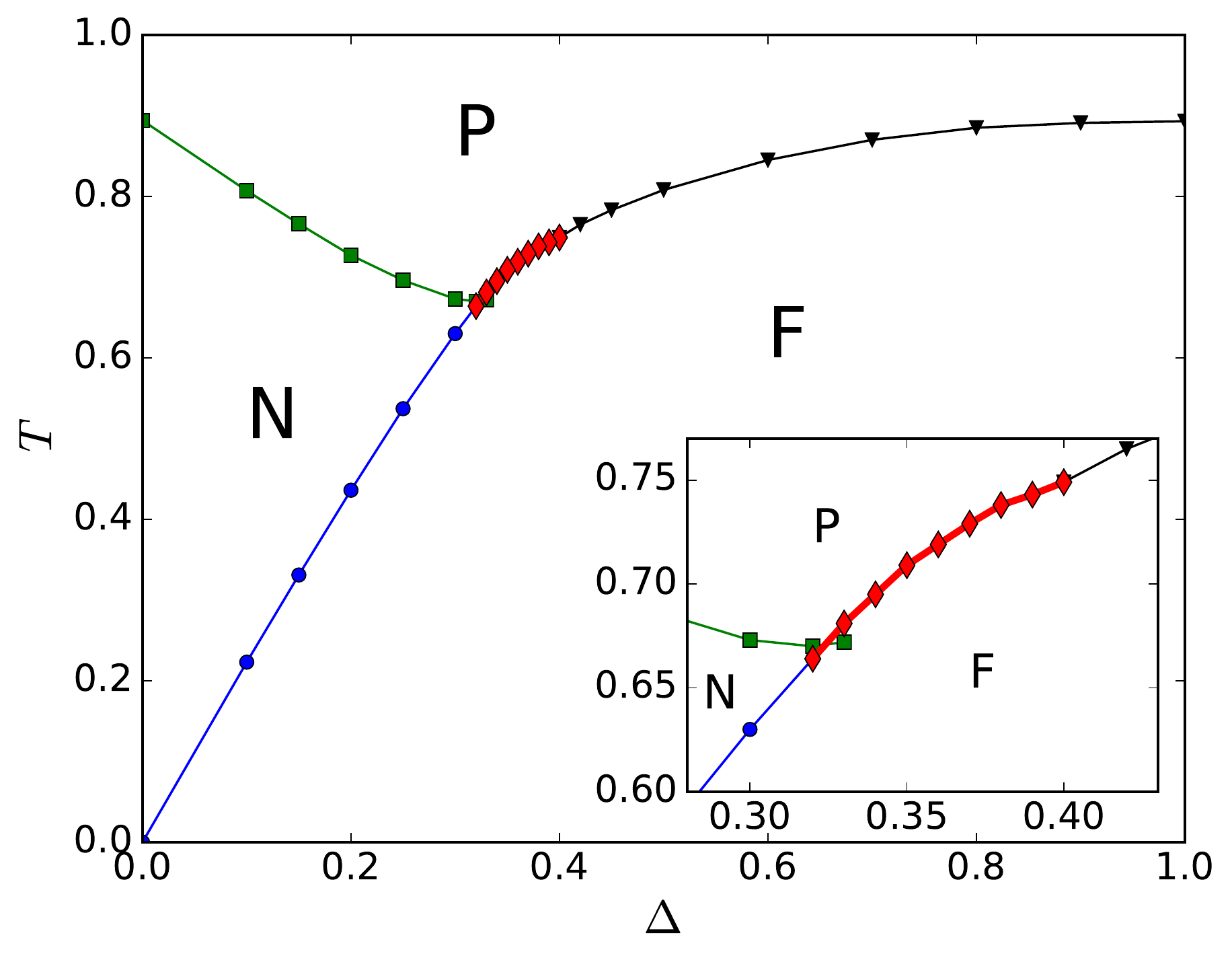}
\caption{\label{fig:phase_diagram}(Color online) The $T$-$\Delta$ phase diagram for the 2D generalized XY model at $q=2$ generated based on the correlation length ratios. The three phases of the diagram are disordered ($P$), nematic ($N$), and quasi-long-range ferromagnetic ($F$). The red diamond thick line is the phase boundary which we focus on in this work. The green square line is obtained based on nematic correlation length ratio. The intersection of these two lines determines the tricritical point $\Delta_c$. The $P\to N$ green square and the $P\to F$ black triangle lines belong to the BKT universality class; the $N\to F$ green circle line belong to the Ising universality class, while the nature of the $P\to F$ red diamond line is under investigation. Inset is the expanded view of the phase diagram in the tricritical region. The error bars are smaller than the symbol sizes.}
\end{figure}

We reproduce the phase diagram of the generalized 2D XY model at $q=2$ [Eq.~\eqref{eq:generalized_xy_ham}] using the correlation length ratios in order to demonstrate the power of these quantities for studying spin models. We find that the correlation length ratios have two distinctive behaviors when the temperature crosses the critical value. In detail, when plotting $\frac{\xi}{L}$ for different $L$'s, they either merge with each other below the critical temperature or cross each other at the critical value. Thanks to these behaviors, we not only determine the critical temperature but also understand the nature of the phase transition. We emphasize that the correlation length is not the optimized method for locating phase transitions with high accuracy, there are better methods for this purpose \cite{Huebscher2013}; instead their main role in this paper is to reveal the physics of the phase transitions.

We present in Fig.~\ref{fig:phase_diagram} the full phase diagram of the model. There are three different phases: (1) the disordered phase ($P$), (2) the quasi-long-range ordered phase ($F$), and (3) the nematic phase ($N$). These phases are characterized by different behaviors of the magnetic and nematic correlation functions as presented in Table~\ref{table:correlation_functions}. Obviously, according to the Mermin-Wagner theorem \cite{Mermin1966}, neither of these phases exhibits long-range order. The three phase boundaries meet at one point, the tricritical point $\Delta_c \approx 0.325$. The result is consistent for the locations of the phases as well as the existence of the tricritical point with previous Monte Carlo studies  which constructed the phase diagram using other physical measurements such as the specific heat and the magnetic susceptibility \cite{Carpenter1989} or the helicity modulus \cite{Huebscher2013}. The interpretation of the correlation length ratios for the construction of the phase diagram is presented below.

\begin{table}[t]
\begin{ruledtabular}
    \begin{tabular}{ccc}
                   & $G_1(r_{ij}) = \langle \cos(\theta_i-\theta_j)\rangle$ & $G_2(r_{ij}) = \langle \cos(2\theta_i-2\theta_j)\rangle$ \\
        \hline 
               $P$ & $\sim\exp[-r_{ij}/\xi_1(T)]$ & $\sim\exp[-r_{ij}/\xi_2(T)]$ \\
               $N$ & $\sim\exp[-r_{ij}/\xi_1(T)]$ & $\sim r^{-\eta_2(T)}$ \\
               $F$ & $\sim r^{-\eta_1(T)}$         & $\sim r^{-\eta_2(T)}$ \\

    \end{tabular}
    \end{ruledtabular}
    \caption{\label{table:correlation_functions} The behaviors of the magnetic [$G_1(r)$] and nematic [$G_2(r)$] correlation functions corresponding to the disordered phase $P$, the nematic phase $N$, and the quasi-long-range ordered phase $F$ \cite{Korshunov1985,Carpenter1989,Huebscher2013}.} 
\end{table}

\subsection{Away from the tricritical point}

In the limit of $\Delta\to 1$ (or $\Delta\to 0$), the model becomes (or equivalent to) the conventional XY model, where there exists only a BKT phase transition from the disordered phase to the quasi-long-range order at $T\approx 0.89$ \cite{Weber1988}. We first discuss the behaviors of the correlation length ratios in proximity to these limits, where the finite size effect is less severe, thus the physics is revealed even at small lattice sizes.

At small $\Delta$ ($\Delta < \Delta_c$), the nematic interaction plays an important role. There are two phase transitions: (1) the $P\to N$ BKT transition at high temperature which is related to the binding/unbinding of half-vortices, and (2) the $N\to F$ Ising transition at a lower temperature occurred when the tension of strings connecting half-vortices vanishes \cite{Korshunov1985,Lee1985,Serna2017} (see Fig.~\ref{fig:phase_diagram}). We will show that, in this range of $\Delta$, $\frac{\xi_n}{L}$ exhibits different behaviors at these two phase transitions, which are attributed to different universality classes. We choose $\Delta=0.2$, which is far from the tricritical point, to examine thoroughly (although as will be discussed later, at $\Delta<\Delta_c$ the finite size effect is not severe).

\begin{figure}[t]
 \includegraphics[width=\columnwidth]{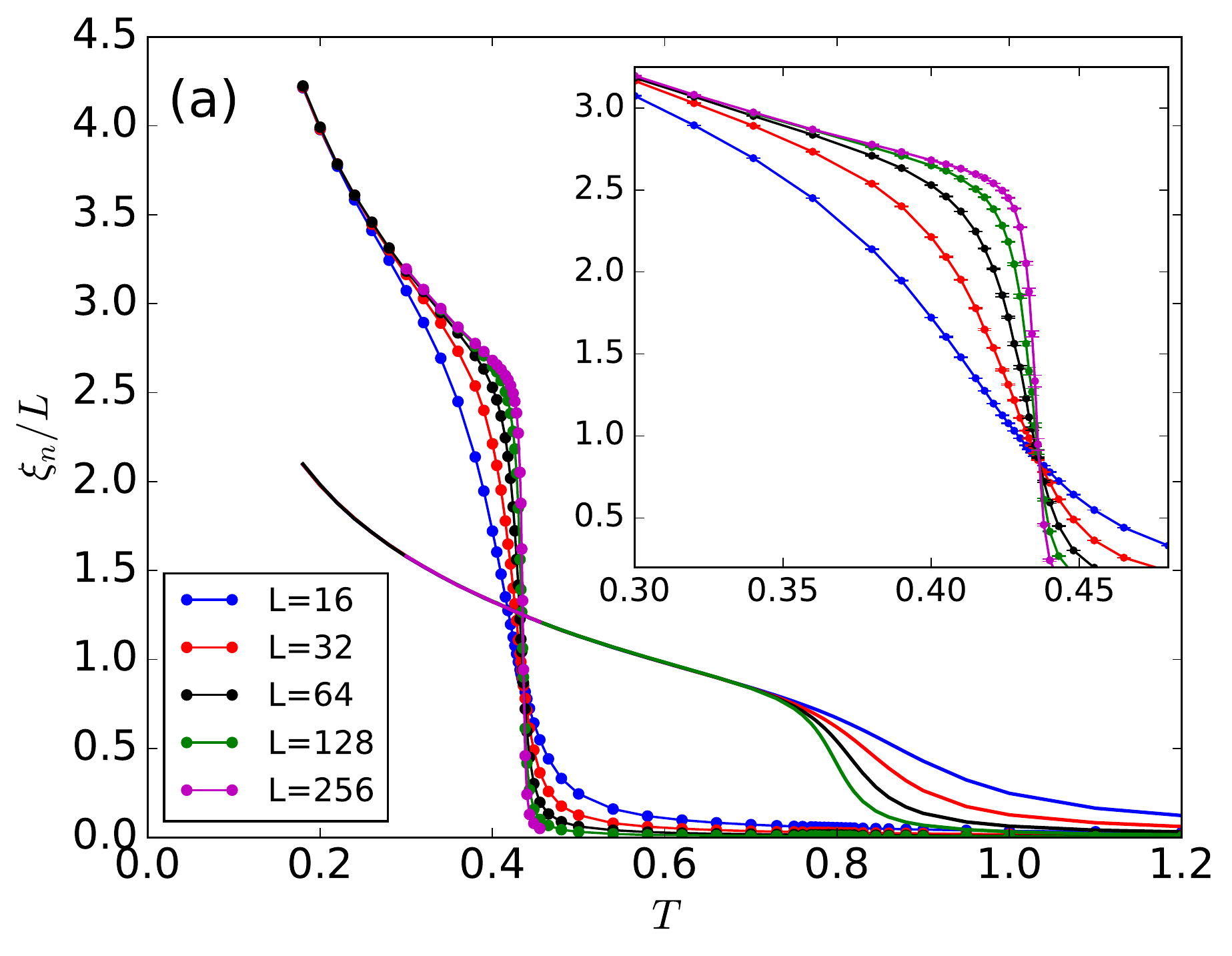}
 
 \includegraphics[width=\columnwidth]{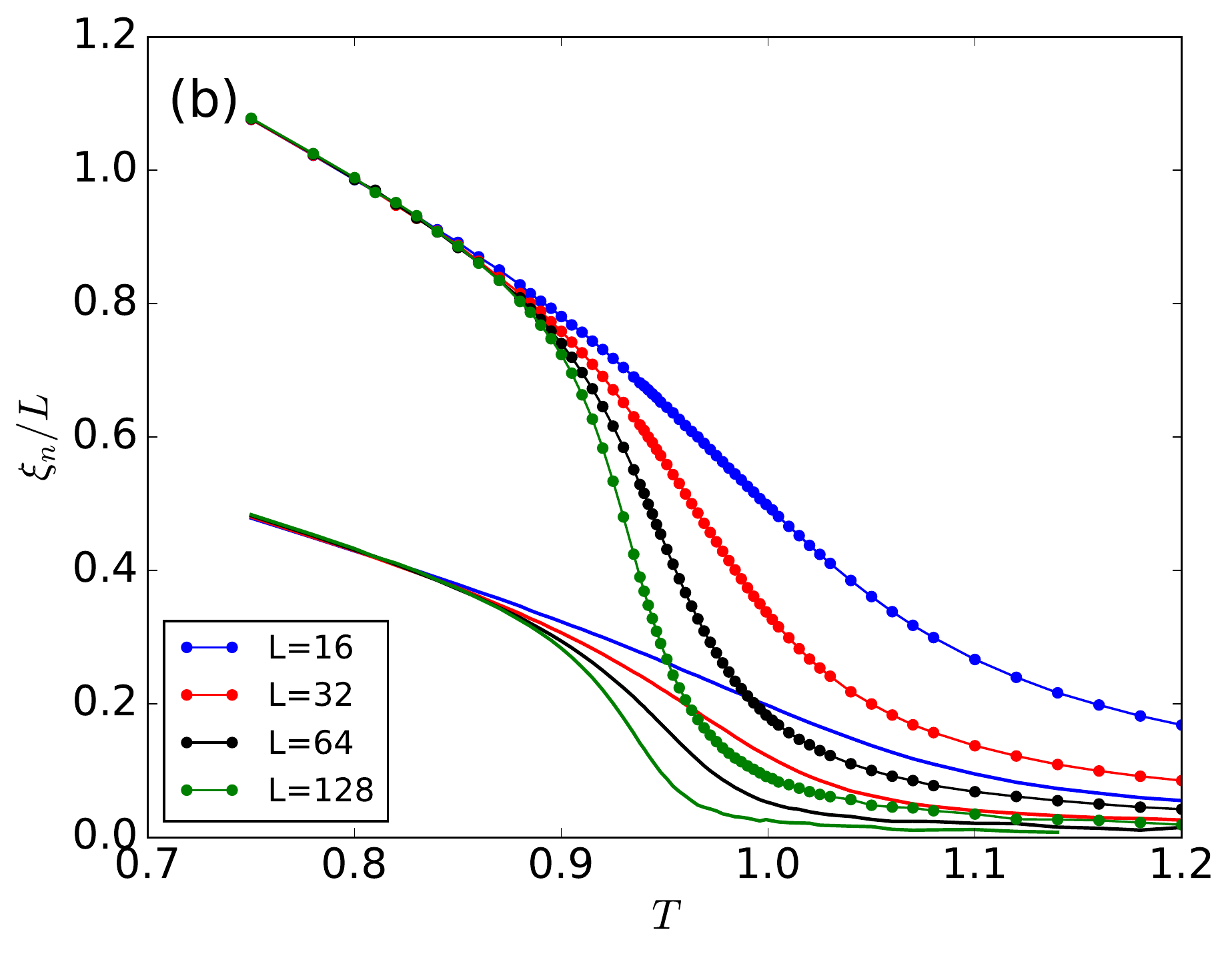}
\caption{\label{fig:easy_cases} (Color online) Temperature-dependent correlation length ratios $\frac{\xi_n}{L}$: (a) at $\Delta=0.2$ (below the tricritical point) and (b) at $\Delta=0.8$ (above the tricritical point). Nematic correlation length ratio: solid lines, magnetic correlation length ratio: solid lines with closed circle symbol. The inset in panel (a) is the expanded view of the magnetic correlation length ratio around and below the crossing temperature. The error bars plotted in the inset of panel (a) are mostly equal or smaller than the symbol sizes.}
\end{figure}

Figure~\ref{fig:easy_cases}(a) shows the correlation length ratios at $\Delta=0.2$. As the correlation length diverges at the critical point, the rapid increases of the correlation length ratio around $T\sim 0.43$ (magnetic curves) and from $0.75$ to $1$ (nematic curves) signal the phase transitions. The magnetic correlation length ratio $\frac{\xi_1}{L}$ (thin solid lines with dot symbols) shows a rapid change at $T\sim 0.43$. The curves of different $L$'s cross at nearly the same point, analogous to the behavior of the Binder parameter $g$ in the Ising model. Finite-size-scaling analysis \cite{Binder1981} shows that in the Ising model the Binder parameter at criticality is universal; the Binder curves at different $L$'s should cross at the critical point if the finite-size correction is absent. Near criticality, the Binder parameter is a function of $\frac{\xi}{L}$ \cite{Binder1981}, thus $\frac{\xi}{L}$ is universal at criticality \cite{Salas2000} and the crossing behavior of the correlation length ratio also means that there is an Ising phase transition with $T_c$ specified in terms of the crossing points (detailed discussion can be seen in Ref.~\onlinecite{Ballesteros2000}). Therefore, in this generalized XY model, the crossing behavior of the magnetic correlation length ratio suggests an Ising phase transition, consistent with previous studies \cite{Carpenter1989,Shi2011,Serna2017}. By extrapolating the crossing point to the thermodynamic limit [see Appendix~\ref{app:therm_limit}], we obtain $T_{c1}=0.436$ for this $\Delta$.

We note that the Ising transition in this model is from the nematic phase to the quasi-long-range ordered phase, which is different from that of the Ising model where it is from the paramagnetic phase to the long-range ferromagnetic order. Therefore, the splaying out of $\xi_1/L$ in Fig.~\ref{fig:easy_cases}(a) occurs only in proximity to the critical temperature. At lower temperature, $\xi_1/L$ curves merge together, characterizing the quasi-long-range order where the system is critical for $T<T_c$, thus $\xi_1/L$ quickly converges to a finite value independent of $L$. The starting temperature for this merging behavior of two $\xi_1/L$ curves increases to the critical temperature as $L$ increases. For example, for the case of Fig.~\ref{fig:easy_cases}(a) we have estimated that the merging point increases from $\sim 0.20$ (for $L=16$ and $32$), $\sim 0.28$ (for $L=32$ and $64$), $\sim 0.30$ (for $L=64$ and $128$) to $\sim 0.36$ (for $L=128$ and $256$). However, as long as $L$ is finite, the crossing behavior for the Ising transition still occurs around $T_c$.

The magnetic correlation length is associated with the correlation between directions of two spins, it is useful to detect if the system is in the quasi-long-range order where the spins tend to align in the same direction for distances smaller than the magnetic correlation length. At higher temperature, it remains finite and does not show any peculiar feature up to $L=256$ [see Fig.~\ref{fig:easy_cases}(a)], although we know that there is $P\to N$ transition at high temperature \cite{Carpenter1989,Huebscher2013}. The reason is that in the nematic case, the spins are only aligned in orientation, not necessary in direction. Hence the nematic correlation length ratio, which only takes care of correlation between orientations of two spins, should be used instead to detect the phase transition at high temperature. Its curves (solid lines without symbols) in Fig.~\ref{fig:easy_cases}(a) shows the signature of a phase transition: $\xi_2$ starts diverging around $0.8$, while $\frac{\xi_2}{L}$ curves of different $L$'s merge with each other, starting in the range of $0.7 < T < 0.9$, thus the critical temperature $T_{c2}$ should be in this range. By extrapolating to the thermodynamic limit (Appendix~\ref{app:therm_limit}), we obtain $T_{c2}\approx0.727$. The merging instead of crossing behavior of $\frac{\xi_2}{L}$ at high temperature suggests that this phase transition be not an Ising transition. We know from previous works that the $P\to N$ phase transition is associated with the binding/unbinding of the half-vortices \cite{Korshunov1985,Lee1985,Huebscher2013}, similar to the transition in the original XY model, it belongs to the BKT universality class. Therefore, we assign the merging behavior of the correlation length ratio to the BKT phase transition (similar finding is found in Ref.~\onlinecite{Ballesteros2000}).

We note that the nematic correlation length ratio does not show any pronounced feature around $T_{c1}$. In this range of temperature, the spins are maintained in nearly the same orientation, while they tend to be in the same senses as the temperature decreases. This is a consequence of the reduction of the lengths of strings connecting half-vortices and then the binding of half-vortices into integer vortices at the phase transition, reflected only in the magnetic correlation length. The nematic correlation length is only related to the orientation of spins; it does not show any feature around $T_{c1}$. Therefore we summarize that (1) depending on the physics of the phases under consideration, we choose an appropriate correlation length to examine and (2) its merging or crossing behavior can determine if the phase transition is of BKT or Ising type, respectively.

With this knowledge in hand, we investigate the region of $\Delta$ close to unity. In this region, the ferromagnetic interaction is dominant; the system is similar to the conventional two-dimensional XY model. For the same reason, we choose the case of $\Delta=0.8$ to analyze thoroughly as it is close enough to $\Delta=1$, the nematic effect is reduced. Figure~\ref{fig:easy_cases}(b) shows the correlation length ratios $\frac{\xi_n}{L}$ of nematic and magnetic types ($n=1, 2$) respectively at $\Delta=0.8$ for different lattice sizes. The two correlation length ratios exhibit the same behavior; they increase as the temperature decreases and merge with each other when $T$ is low enough. The rapid increase of both $\frac{\xi_n}{L}$ at $T$ around $0.85\to 1$ implies that the critical temperature is in this range. Based on the merging behavior, we conclude that there is a BKT phase transition, and by extrapolation to $L\to\infty$ we obtain $T_c=0.885$. Indeed, this is a phase transition from the disordered phase where the spins are set randomly to the quasi-long-range order where the spins are aligned both in the sense and orientation. Therefore both nematic and ferromagnetic correlation lengths are sensitive to this phase transition, explaining the merging behavior of both quantities. One can use the two correlation lengths interchangeably to detect the phase transition.

\subsection{Tricritical region}

\begin{figure}[t]
 \includegraphics[width=\columnwidth]{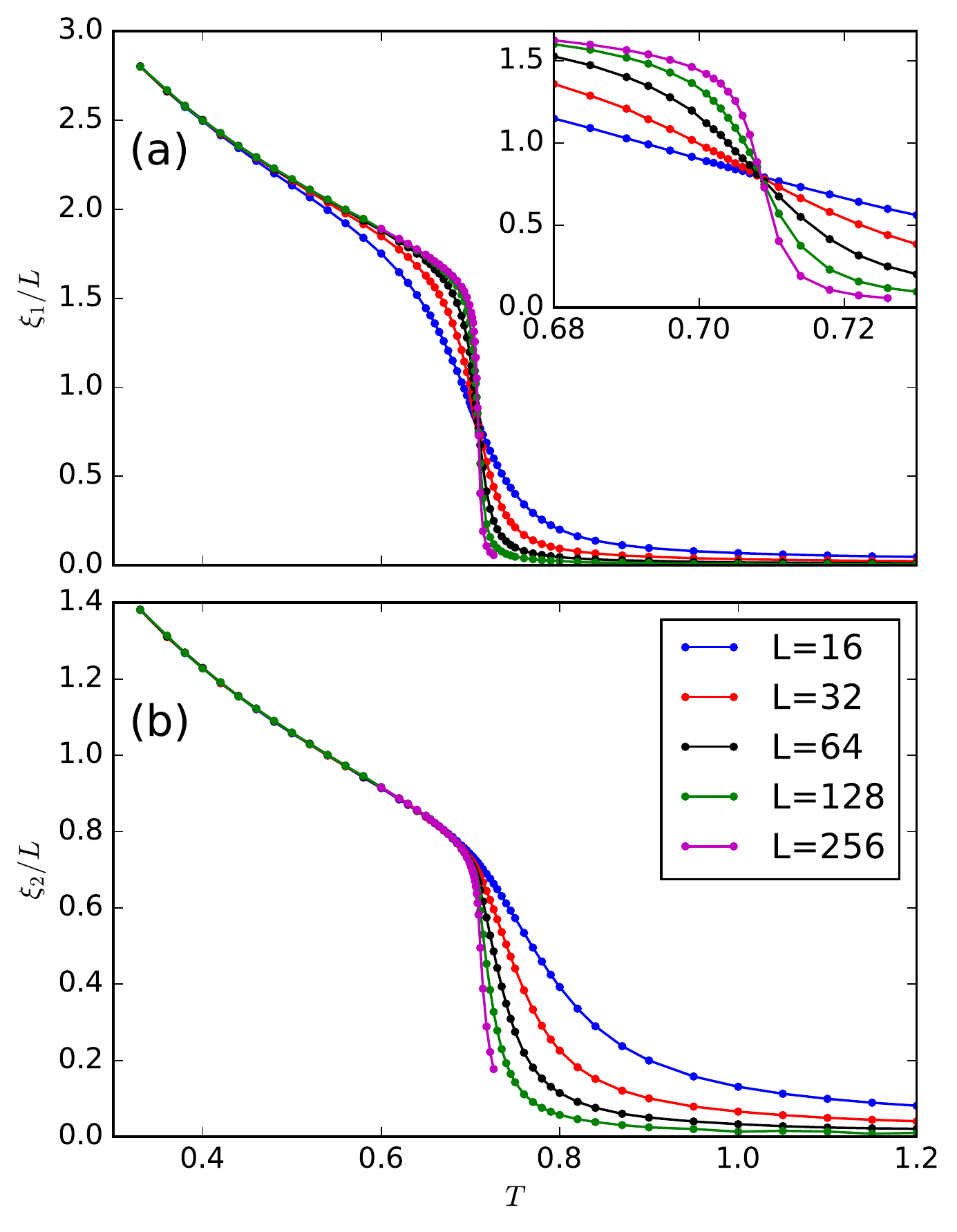}
\caption{\label{fig:near_critical} (Color online) Temperature-dependent correlation length ratios $\frac{\xi_n}{L}$ at $\Delta=0.35$ (slightly above the tricritical point $\Delta_c\approx0.325$): (a) the magnetic correlation length ratio, (b) the nematic correlation length ratio. Inset in panel (a) is the expanded view of the magnetic correlation length ratio around the crossing point. The error bars are equal or smaller than the symbol sizes.}
\end{figure} 

We focus on the region of the phase diagram where both the nematic and magnetic terms contribute significantly to the physics of the system. This region is specified by $\Delta$ away from $0$ and $1$, and mostly in the range around the tricritical point $\Delta_c\approx0.325$. The difficulty arising in this region is that due to the competition between the nematic and magnetic interactions, the correlation length ratios at finite sizes may contain features from the phases below and above $\Delta_c$. Our goal is thus to determine the phase transitions occurred in this region from this mix of features. For that purpose, we investigate the cases $\Delta=0.3$ and $0.35$, which are slightly below and above the tricritical point $\Delta_c$.

Interestingly, the case $\Delta=0.3$ is qualitatively similar to the case $\Delta=0.2$ discussed previously, despite its proximity to the tricritical point. The correlation length ratios (not shown) behave in the same manner as those at $\Delta=0.2$. For example, there are crossing points at $T_{c1}$ for $N\to F$ transition in the magnetic correlation length ratio, suggesting an Ising phase transition; for $P\to N$ phase transition, $\frac{\xi_2}{L}$ curves of different $L$'s start merging at $T_{c2}$ toward lower temperature, marking a BKT transition. As will be shown in the next section, the crossing behavior tends to be stable in the thermodynamic limit. Thus we have not found any effect of the BKT transition from above the tricritical point in the $N\to F$ phase transition. As a result, for $\Delta < \Delta_c$, the critical temperatures are obtained straighforwardly.

In contrast, at $\Delta=0.35$ which is slightly above $\Delta_c$, there is influence of the nematic phase on $\frac{\xi_n}{L}$. Figure~\ref{fig:near_critical} shows the plots of both $\frac{\xi_n}{L}$ as functions of temperature for $\Delta=0.35$. On the one hand, the magnetic correlation length ratio in Fig.~\ref{fig:near_critical}(a) still exhibits the crossing behavior with a large change at the critical temperature, similar to that of $\Delta < \Delta_c$. However, it is not clear if the curves merge together when $L$ increases. Assuming the crossing behavior, we specify $T_{c1}\approx 0.709$. On the other hand, the nematic correlation length ratio retains the merging feature [Fig.~\ref{fig:near_critical}(b)]. It is reasonable as it characterizes the change from the paramagnetic to the two quasi-long-range orders (the $N$ and $F$ phases), where the spins in an arbitrary domain tend to be at least in similar orientation. Based on the merging behavior of the nematic correlation length ratio, we obtain $T_{c2} \approx 0.692$. Thus $T_{c1}$ and $T_{c2}$ are not numerically the same, the difference $0.017$ is mostly due to our ways to extrapolate $T_c$, especially for $T_{c2}$.

To specify $\Delta_c$, we compare the transition temperature $T_{c1}$ and $T_{c2}$ obtained from magnetic and nematic correlation length ratio, respectively, for a range of $\Delta$ in proximity to the tricritical point. For $\Delta < \Delta_c$, that $T_{c2} > T_{c1}$ corresponds to two phase transitions $P\to N$ and $N\to F$. For $\Delta > \Delta_c$, theoretically $T_{c1} = T_{c2}$, but our measurements exhibit that systematically $T_{c1}$ is slightly larger than $T_{c2}$ by less than $0.02$ near $\Delta_c$ , thus it is the uncertainty of the method. Nevertheless, $\Delta_c$ satisfies $T_{c1} = T_{c2}$. By interpolation, $\Delta_c\approx 0.325$ is obtained (see the inset of Fig.~\ref{fig:phase_diagram}).

Therefore, by measuring both the magnetic and nematic correlation lengths, we reconstruct the full phase diagram of the generalized two-dimensional XY model at $q=2$ as in Fig.~\ref{fig:phase_diagram}. The result is consistent with previous Monte Carlo studies for the same model \cite{Carpenter1989,Huebscher2013}. The only mysterious point is the physics slightly above the tricritical point, where there may exist both crossing and merging behaviors in the correlation length ratios for the same phase transition. This is the main reason for our study and we devote the next section for the understanding of this issue.

\section{Ising transition line \label{sec:ising_transition}}

\begin{figure}[t]
 \includegraphics[width=\columnwidth]{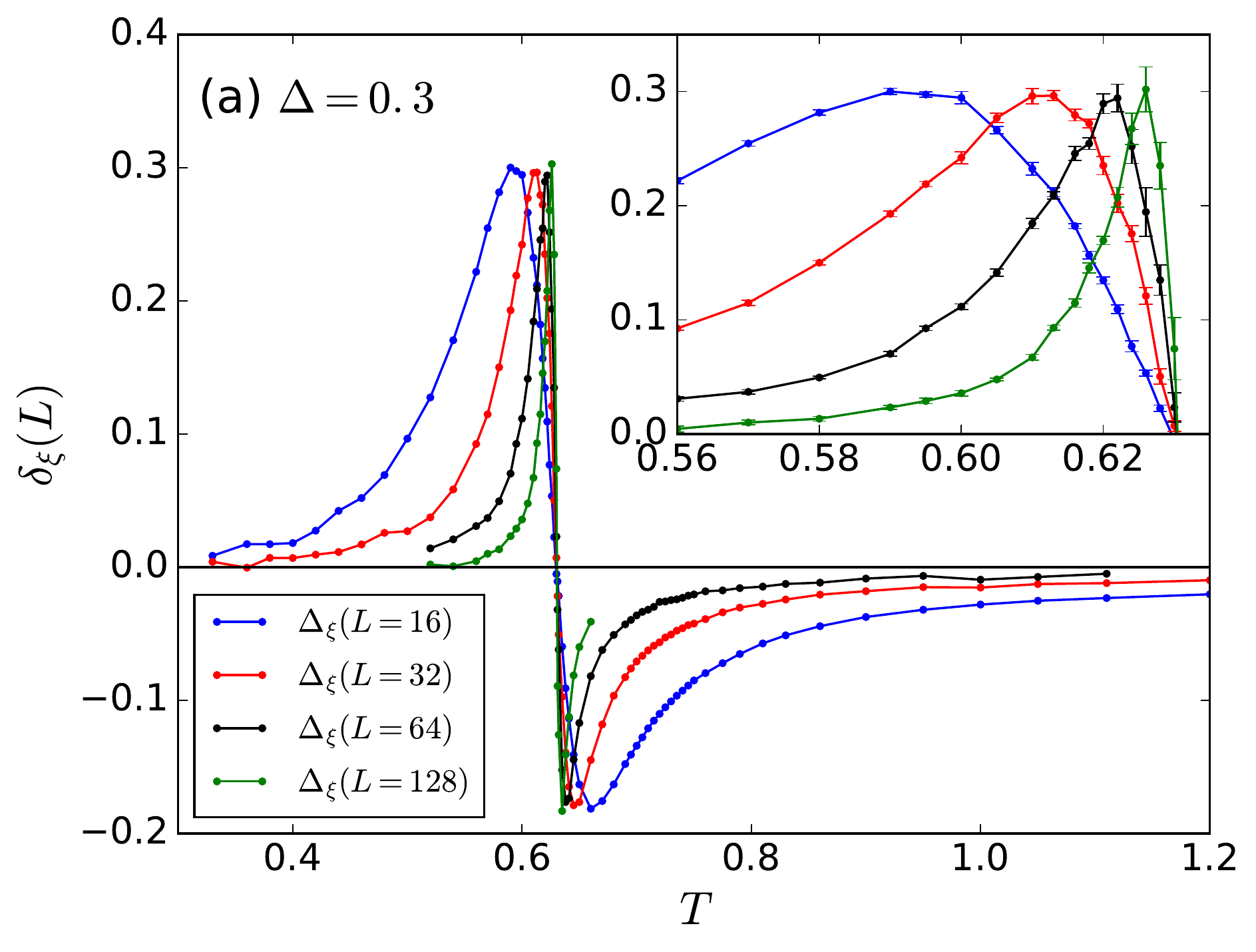}
 
 \includegraphics[width=\columnwidth]{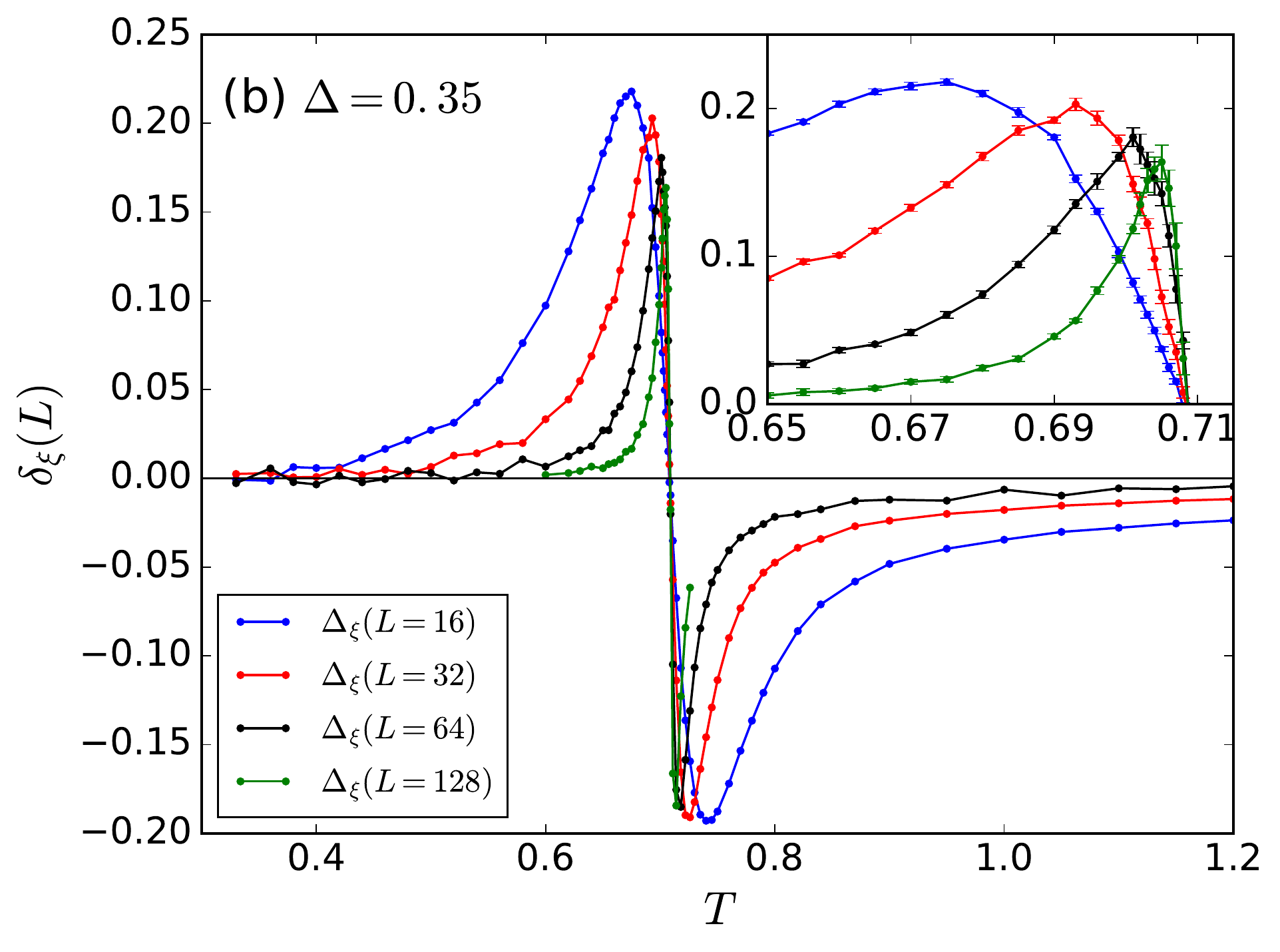}
\caption{\label{fig:diff_corr_length} (Color online) Temperature dependence of the difference of the magnetic correlation length ratio $\delta_\xi(L)$ [Eq.~\eqref{eq:delta_corr}]. Panel (a): $\Delta=0.3 < \Delta_c$. Panel (b): $\Delta=0.35 > \Delta_c$. ($\Delta_c\approx0.325$.) Insets are the expanded views of the differences in the region below $T_c$ where they reach maximal values. Typical error bars are shown in the insets.}
\end{figure}

The consensus is that the phase transition from the nematic to the quasi-long-range order below the tricritical point (see Fig.~\ref{fig:phase_diagram}) belongs to the Ising universality class \cite{Carpenter1989,Huebscher2013,Shi2011,Serna2017}. The remaining open questions focus on the tricritical region: (1) whether this Ising line goes beyond the tricritical point, (2) the nature of the Ising segment beyond the tricritical point if there is and (3) the transition from the Ising to the BKT universality class along this line. Some of these issues have been studied in Refs.~\onlinecite{Shi2011,Serna2017}, which focus on the modified Villain model of Eq.~\eqref{eq:generalized_xy_ham} and claim that the Ising transition line goes beyond the tricritical point. The Monte Carlo study of Ref.~\onlinecite{Huebscher2013} directly simulates Eq.~\eqref{eq:generalized_xy_ham} but only briefly mentions the possibility of the Ising transition above the tricritical point based on the specific heat measurements. In Appendix~\ref{app:specific_heat}, we reexamine the specific heat carefully and find that while the result is consistent with Ref.~\onlinecite{Huebscher2013}, there is other information unable to observe in the specific heat measurements and in other previous studies. By using the correlation length ratios, we hope to give more insights to some of the above questions.

\begin{figure}[t]
 \includegraphics[width=\columnwidth]{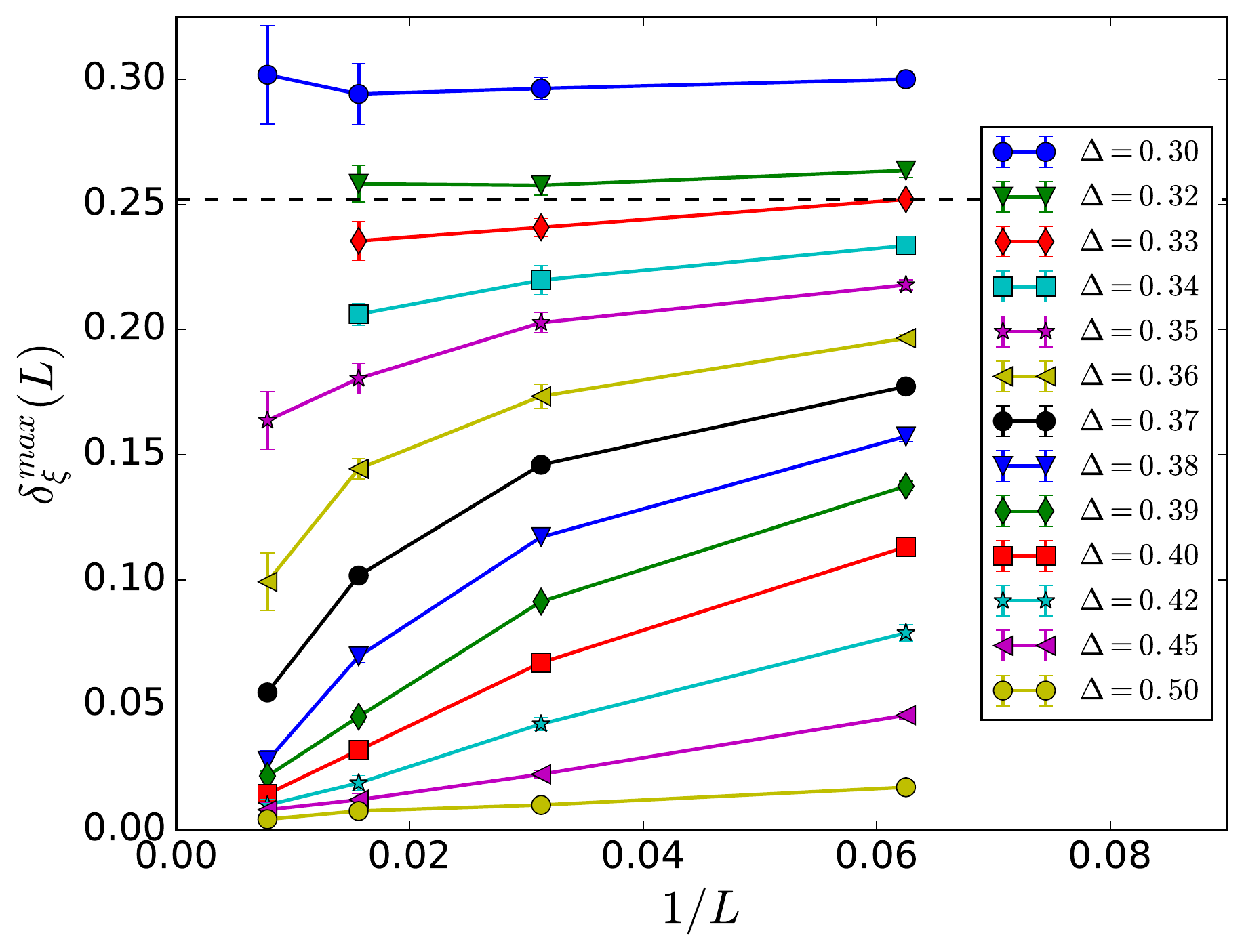}
\caption{\label{fig:xi_L_difference} (Color online) Maximum of the difference in $\frac{\xi_1}{L}$ as a function of $\frac{1}{L}$ for a wide range of $\Delta$. The dashed horizontal line separates two regions: $\Delta < \Delta_c$ above the line, and $\Delta > \Delta_c$ below the line.}
\end{figure}

First, we note that there is ambiguity if observing directly the correlation length ratios. As discussed in Sec.~\ref{sec:phase_diagram}, in the region of $\Delta \gtrsim \Delta_c$ (such as $\Delta=0.35$), the magnetic correlation length expresses the crossing behavior. Due to the limit of the computational resource, one may not be able to answer whether the crossing behavior changes to merging behavior at larger lattice sizes or it is maintained in the thermodynamic limit. In contrast, the nematic correlation length ratio clearly shows the merging behavior, a signature of the BKT-type transition, but this $\frac{\xi_2}{L}$ relates more to the pairing of half-vortices. Therefore, it is not conclusive yet if the Ising line goes beyond the tricritical point from this view.

We find that it is much easier to speculate the physics at large $L$ by studying the difference of the magnetic correlation length ratio at different $L$ values only with data for a small set of finite $L$'s. We define the difference as
\begin{equation}\label{eq:delta_corr}
 \delta_\xi(L)=\dfrac{\xi_1(2L)}{2L} - \dfrac{\xi_1(L)}{L}.
\end{equation}
Figure~\ref{fig:diff_corr_length} shows the difference for $\Delta=0.3$ and $0.35$, the narrow range where the critical value $\Delta_c\approx0.325$ is in between. The trends for $\Delta$ are pronounced. Let $\delta^{max}_\xi(L)$ be the maximum of the difference $\delta_\xi(L)$ for $T < T_c$. For $\Delta=0.3 < \Delta_c$ (which is known to exhibit the Ising phase transition), below the crossing point, $\delta^{max}_\xi(L)$ are constant with respect to the lattice size [inset of Fig.~\ref{fig:diff_corr_length}(a)]. It suggests that $\delta^{max}_\xi(L)$ be unchanged for $L\to\infty$, i.e. the crossing behavior is maintained in the thermodynamic limit, confirming that the phase transition is of Ising type. In contrast for $\Delta=0.35 > \Delta_c$, the maxima below the crossing temperature of the difference decreases systematically. This signal is easily observed even at small lattice sizes, e.g. in Fig.~\ref{fig:diff_corr_length}(b), with $L$ running from $16$ to $128$, one already observes the decreasing tendency of the maxima. As a result, the phase transition at $\Delta=0.35$ is not of the same type as the $N\to F$ Ising transition. Depending on $\delta_\xi(L)$ in the thermodynamic limit, there are two possibilities for this $P\to F$ phase transition: (1) the maximum of $\delta_\xi\to 0$ as $L\to\infty$, the $\frac{\xi_1}{L}$ curves change to merging behavior, it is more likely to be a BKT-type transition or (2) the maximum reaches a finite value, the crossing behavior is maintained, then it is another Ising-type transition, but may not have the same physics as that of the $N\to F$ transition.

Figure~\ref{fig:xi_L_difference} is another view of the difference in correlation length ratio. It is the plot of $\delta_\xi(L)$ versus $1/L$, showing the tendency of the maxima of $\delta^{max}_\xi$ as $L$ increases. At $\Delta\le\Delta_c$, the curves are horizontal lines that tend to reach the $L\to\infty$ limit at finite values, confirming that the phase transition from $N\to F$ belongs to the Ising universality. For $\Delta_c < \Delta < 0.4$, the curves bend down, at $\Delta$ close to $\Delta_c$, it requires larger-scale simulations to understand the physics, however, for $\Delta \ge 0.36$, the tendency toward zero can be observed. At $\Delta\ge0.4$, the curves become rather linear and clearly go to zero, thus the merging behavior can occur at large enough $L$ for $\Delta \ge 0.4$, confirming the BKT phase transition. Therefore, the range of interest is $\Delta_c < \Delta < 0.4$, and while we cannot access larger-scale simulation, at least for $\Delta \ge 0.36$, we can say that the phase transition is not truely an Ising-type transition, as the crossing behavior of the correlation length ratio is not maintained in the thermodynamic limit. It is not truely of BKT type either as $\delta_\xi^{max}(L)$ goes to zero rather slowly, thus behaving differently from that at $\Delta \ge 0.4$.

\begin{figure}[t]
 \includegraphics[width=\columnwidth]{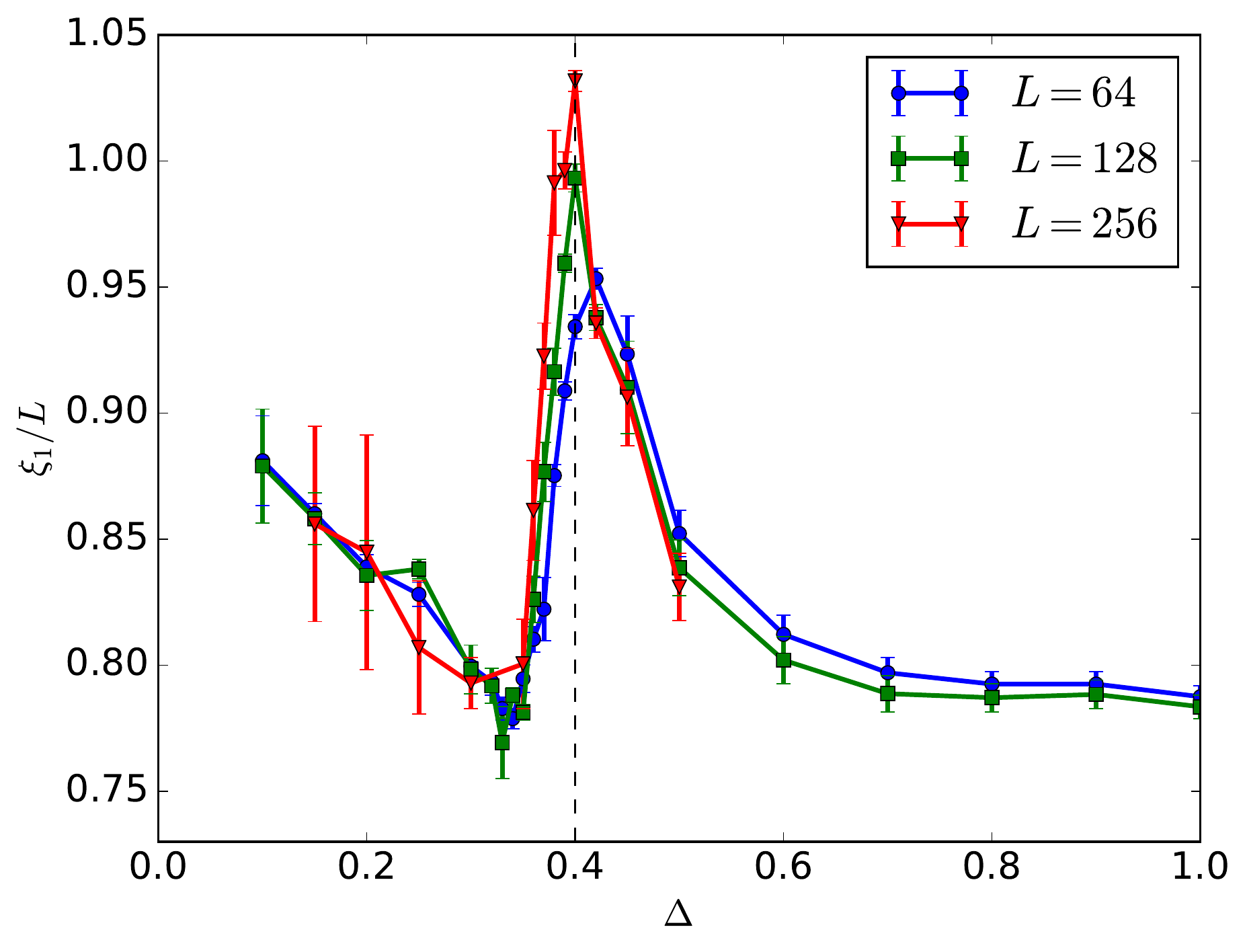}
\caption{\label{fig:xi_L_critical} (Color online) The value of $\frac{\xi_1}{L}$ at the critical temperature $T_c$ plotted vs. $\Delta$ for three cases denoted by $L$ values. For $\Delta \le 0.4$, $\frac{\xi_1}{L}$ for each $L$ is determined by the crossing point of the correlation length ratios obtained from the simulations at sizes $L/2$ and $L$. For $\Delta>0.4$, $\frac{\xi_1}{L}$ for each $L$ is the value of the correlation length ratio from the simulation at size $L$ at $T=T_c$ (the critical temperature at $L\to\infty$). The vertical dashed line marks $\Delta=0.4$.}
\end{figure}

Furthermore, we examine $\frac{\xi_1}{L}$ at the critical temperature. The correlation length ratio at criticality is claimed to be universal \cite{Salas2000,Zierenberg2017}, hence it can be a criterion for classifying the phase transitions in this study. Figure~\ref{fig:xi_L_critical} shows the measurements for a wide range of $\Delta$, except for the point $\Delta=0$ where there is no phase transition from $N$ to $F$. These are obtained directly from our finite-size simulations for different lattice sizes. We do not use the extrapolated values at $L\to\infty$ as the lattice sizes in use are small, thus the extrapolations are not in high quality, especially in the region around the tricritical point. As $\Delta\to 0$, $\frac{\xi_1}{L}$ approaches the critical value for the Ising model ($\sim 0.905$) \cite{Salas2000}. At $\Delta = 1$, which is the original XY model, it is $\sim0.78$, consistent with Refs.~\onlinecite{Hasenbusch2005,Komura2012} and close to the exact value $\approx0.75$ \cite{Hasenbusch2005,Komura2012}. Thus at the two limits $\Delta\to 0$ and $\Delta=1$, the value of the correlation length ratio shows clear signatures that the phase transition is of Ising and BKT type, respectively. However for $\Delta$ away from zero and one, as being weakly universal, the critical value of $\frac{\xi_1}{L}$ strongly depends on $\Delta$. Below $\Delta_c$, it decreases linearly with increasing $\Delta$, while above $\Delta=0.4$, it decreases nonlinearly, characterizing two regions of Ising and BKT-type phase transition, respectively. We note that the large error bars at $\Delta < \Delta_c$ is because the temperature mesh for large-$L$ simulation is not fine enough, thus the crossing point is determined with large uncertainty. In the $\Delta$-range of our interest, the critical $\frac{\xi_1}{L}$ changes abruptly from the minimum at $\Delta\sim \Delta_c$ and reaches the maximum at $\Delta\sim0.4$. The trend of $\frac{\xi_1}{L}$ as $L$ increases depends on $\Delta$: for $\Delta < \Delta_c$, it converges rapidly, for $\Delta>0.4$, it decreases and converges more slowly, while $\Delta_c < \Delta < 0.4$, it increases rather fast. It supports the sudden change of $\frac{\xi_1}{L}$ around the tricritical region. Hence, despite results at finite-size simulations, Fig.~\ref{fig:xi_L_critical} is still physically meaningful, it suggests that there exists a narrow region of $\Delta$ that separates the two different types of phase transition (Ising and BKT types), where certain quantities behave differently or change rapidly. The range $\Delta_c < \Delta < 0.4$ is such a region, characterized by the nonlinear bending of the $\delta_\xi(L)$ and the rapid increase of $\frac{\xi_1}{L}$ at criticality.

Therefore, the region of $\Delta_c < \Delta < 0.4$ is a special one. It may be related to the region for the ``deconfinement phase transition'' proposed by Serna \textit{et al.} \cite{Serna2017}. However the ``deconfinement'' physics is not clear from the perspective of the correlation length. Instead the correlation length ratio can only separate this region from the those of the Ising and BKT transitions. The upper limit $\Delta\approx0.4$ is detected by both the correlation length ratio and the specific heat (see Appendix~\ref{app:specific_heat}), distinguishing it from the usual BKT transition at larger $\Delta$, while the lower limit at $\Delta_c$ can only be observed using the correlation length ratio. However both $\frac{\xi_1}{L}$ at criticality and $\delta_\xi(L)$ show that this phase transition segment is not a continuation of the Ising line of the $N$-$F$ transition, which is unable to observe using other quantities such as the specific heat.  We believe that the nature of the phase transition in this segment is different from that of other segments. It can be considered as the intermediate region connecting the Ising transition line and the BKT transition line. Therefore, in the phase diagram, we distinguish it (the diamond red line in Fig.~\ref{fig:phase_diagram}) from other phase transition lines.

\section{Conclusions \label{sec:conclusions}}

In this paper, we have studied in detail the behaviors of the magnetic and nematic correlation length ratios $\frac{\xi_n}{L}$ in the two-dimensional generalized XY model at $q=2$. We demonstrated the power of $\frac{\xi_n}{L}$ in determining the phase transitions as temperature decreases for a wide range of the nematic interaction strength with respect to the magnetic interaction. We showed how to classify the type of a phase transition based on the behavior of $\frac{\xi_n}{L}$, without directly calculating the critical exponents. More importantly, we investigated the region around the tricritical point $\Delta_c\approx 0.325$ and found pronounced features of $\frac{\xi_n}{L}$ that give insights into the physics of this region.

We have several results. First, the correlation length ratios exhibit different behaviors depending on whether the phase transition is of Ising or BKT type for this generalized XY model. For Ising phase transitions, based on the finite-size scaling argument \cite{Binder1981}, magnetic curves $\frac{\xi_1}{L}$ for different lattice sizes cross at the critical temperature. For BKT phase transitions, both $\frac{\xi_n}{L}$ curves merge together at lower temperature, the merging point approaches the critical point as the lattice size increases. The correlation length ratios appear to be sensitive to phase transitions even with small lattice sizes ($L\le 256$ as in this study), thus it is useful when simulations for large lattice sizes cannot be accessed.

Second, the observations of the magnetic correlation length ratios at critical temperature and its difference between lattice sizes $L$ and $2L$ show that the phase transition in the range from $\Delta_c\approx0.325$ to $0.4$ exhibits different physics from those below $\Delta_c$ or above $0.4$. The Ising line does not connect directly to the BKT line in the phase diagram. Instead the region from $\Delta_c$ to $0.4$ plays the role of an intermediate region where the behaviors of the related quantities (e.g. the correlation length) change from more Ising-like near $\Delta_c$ to more BKT-like near $\Delta\approx 0.4$. The phase transition in this region is however neither of Ising type nor BKT type. 

Our study contains limitations. First, due to the limit of our computational resource, we can only carry out simulations with the lattice size as large as $L=256$. With larger $L$, such as $L=512$ and $1024$, and with better Monte Carlo statistics, which are feasible at present (given enough computational resource), one could observe more clearly the behavior of $\frac{\xi_1}{L}$ in proximity to the tricritical point, and extrapolate more accurately critical values in the thermodynamic limit. Simulations at larger scale may be helpful to confirm the behavior at $\Delta$ very close to $\Delta_c$ where the ``critical slowing down'' is severe. Second, the correlation length is not the method of choice to locate the phase boundary with accuracy; other measurements such as the helicity modulus \cite{Huebscher2013} perform better. The role of the correlation length ratio is to understand the physics of the phase transitions in this model; the construction of the full phase diagram with accuracy is not the aim of this work.

Finally, we indicate several problems arising from the applications of $\frac{\xi_n}{L}$. First, the dependence of $\Delta$ on the critical value of $\frac{\xi_1}{L}$ or the dependence on $L$ of its difference have been investigated for small lattice sizes. Simulations at larger scales would be necessary to confirm the behavior of the correlation length in the tricritical region. More importantly, the phase transition in this region has only been found to be neither of Ising type nor BKT type, its nature, whether it is the ``deconfinement transition'' as in Ref.~\onlinecite{Serna2017} and how one relates to this ``deconfinement'',  is however not fully understood. These are open questions for future study.

\section*{Acknowledgments}
We are grateful to Stefan Wessel, Junichi Okamoto and Mai Suan Li for useful discussion. This research is funded by Vietnam National Foundation for Science and Technology Development (NAFOSTED) under Grant No. 103.02-2011.38. Numerical calculation was performed at the Advanced Institute for Science and Technology, Hanoi University of Science and Technology.

\appendix

\section{Specific heat \label{app:specific_heat}}

\begin{figure}[t]
 \includegraphics[width=\columnwidth]{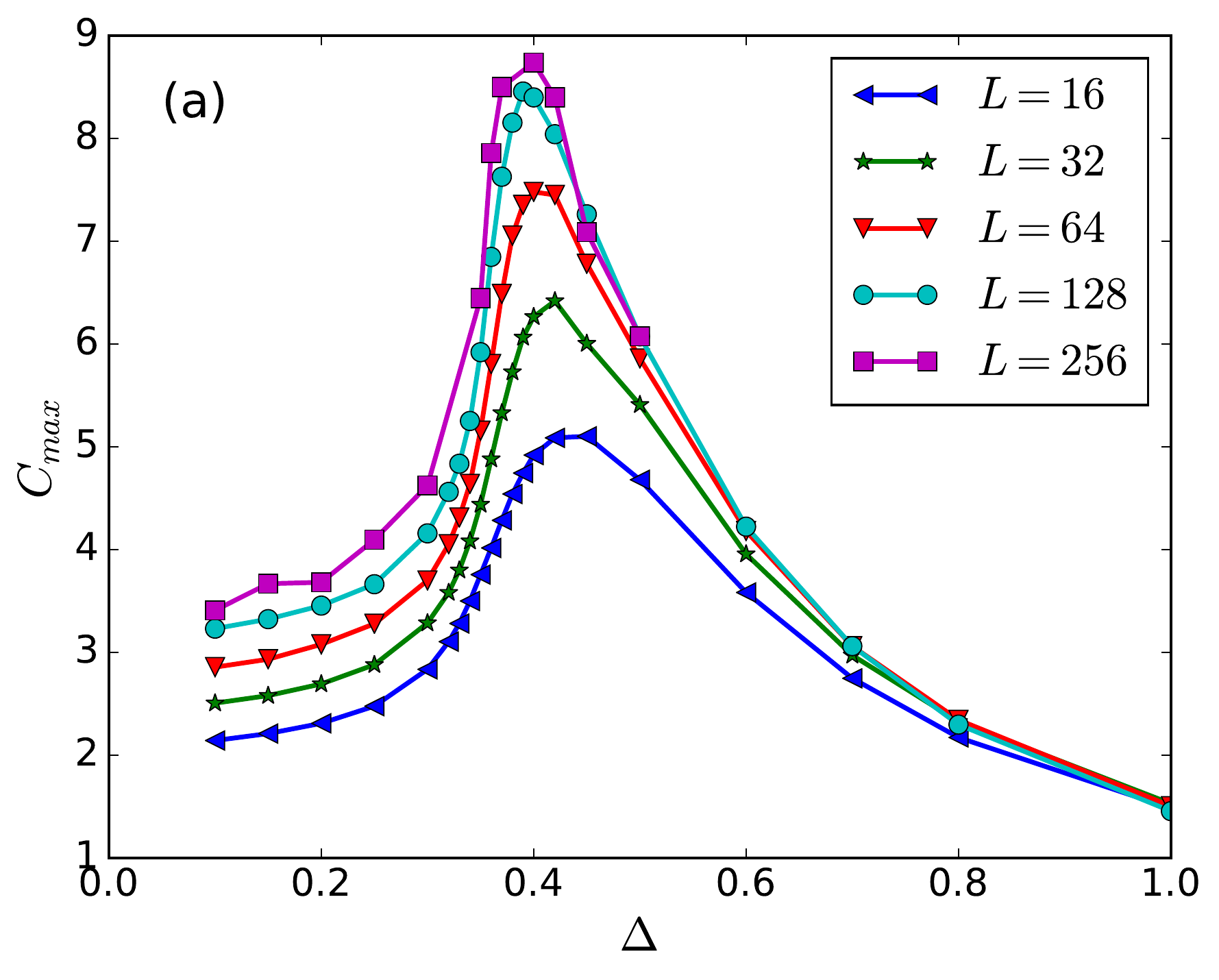}
 
 \includegraphics[width=\columnwidth]{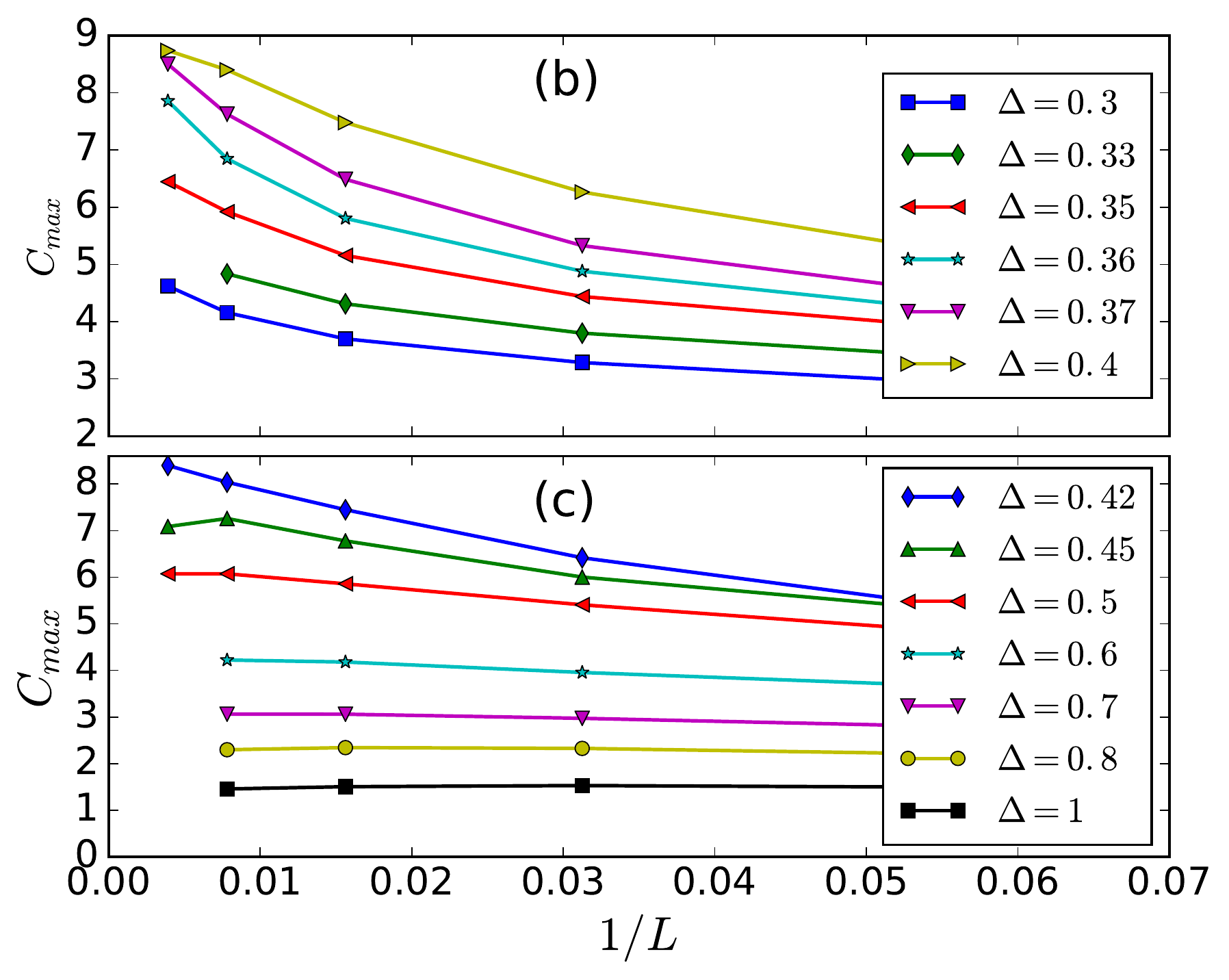}
\caption{\label{fig:specific_heat} (Color online) (a) The plot of the specific heat maximum $C^{max}_V$ as a function of $\Delta$, exhibiting a peak at $\Delta=0.4$. (b), (c) $C^{max}_V$ as a function of $1/L$ for $\Delta \le 0.4$ [below the peak position - panel (b)] and for $\Delta > 0.4$ [above the peak position - panel (c)]. The error bars are smaller than the symbol sizes.}
\end{figure}

From the viewpoint of Monte Carlo simulation, H\"ubscher and Wessel \cite{Huebscher2013} studied the maxima of the specific heat ($C_{max}$) and found that the specific-heat-maximum line approaches the BKT transition line slightly above the tricritical point. As the shapes of the specific heat peaks may be used to determine the type of the phase transition \cite{Carpenter1989}, they suggested that the Ising line might be dominant in the region of the critical point.

Here we reconsider the maximal value of the specific heat and plot in Fig.~\ref{fig:specific_heat} the magnitude $C_{max}$ obtained from our calculations. We note that while at $\Delta < \Delta_c$, there are two peaks for $C$ at different temperatures, corresponding to the $N$-$F$ and $P$-$N$ phase transition, we only focus on the peak for $N$-$F$ transition at low temperature. We do not examine the temperature at which $C$ is maximized, as it has been already considered \cite{Huebscher2013}, instead we investigate the dependence of the $C_{max}$ value on $\Delta$ and $L$. In Fig.~\ref{fig:specific_heat}(a), $C_{max}$ is plotted as a function of $\Delta$ which reaches the maximum at $\Delta\approx0.4$ for all $L$ values that we have calculated. The value $\Delta=0.4$ separates two regions: (1) the region $\Delta \le 0.4$, where the specific heat peaks are sharp and increases as $L$ increases, is associated with the Ising transition, (2) the region $\Delta > 0.4$, where the peaks are smaller and broader (usually located at temperature above $T_c$), is in connection with the BKT transition. If plotting $C_{max}$ against $L$, the concavity of the curves changes at $\Delta\approx 0.4$ as illustrated in  Figs.~\ref{fig:specific_heat}(b) and (c). Figure~\ref{fig:specific_heat}(c) for $\Delta>0.4$ (except for the case $\Delta=0.42$ which requires simulations at larger $L$) shows the convergence of $C_{max}$ to finite values as $L$ is large enough, signifying the BKT transition. For $\Delta\le 0.4$ in Fig.~\ref{fig:specific_heat}(b), $C_{max}$ keeps increasing as $L$ increases, implying divergence at $T\to T_c$, suggesting the Ising transition.

Therefore Fig.~\ref{fig:specific_heat} exhibits the characteristics of the Ising transition up to $\Delta\sim 0.4$, consistent with Ref.~\onlinecite{Huebscher2013}. On one hand, the fact that $C_{max}$ reaches maximal at $\Delta\approx 0.4$ supports the $\frac{\xi_1}{L}$ results in the main text for the change to BKT phase transition at this $\Delta$. On the other hand, the specific heat does not show any pronounced feature at the tricritical point $\Delta_c\approx 0.325$, instead it predicts the Ising-like transition for the whole range $\Delta<0.4$, while the correlation length shows critical behaviors at $\Delta_c$. It means that there is other physics not revealed by the specific heat. We believe that at this $\Delta$, the total energy of the system is varied smoothly, thus there is no peculiarity from the observation of the specific heat. The pronounced feature of the correlation length at $\Delta_c$ means that there are topological changes which are irrelevant to the total energy.

\section{Thermodynamic limit \label{app:therm_limit}}

To extrapolate the critical temperatures $T_c$ for the phase transitions in the thermodynamic limit, which are then used for the phase diagram (Fig.~\ref{fig:phase_diagram}), we apply two different ways for the Ising and BKT transition, respectively. For the Ising transition, similar to the Binder parameter analysis \cite{Binder1981}, as long as the magnetic correlation length ratio curves cross with each other, we specify the crossing temperature $T_c(L)$ for each pair of size $L$ and $2L$, the limit at $L\to\infty$ is obtained by linearly fitting $T_c(L)$ vs. $1/L$. We employ this procedure to determine $T_c$ for $\Delta < 0.4$.

For the BKT phase transition, we apply the method from Ref.~\onlinecite{Surungan2005}. For the set of $\frac{\xi_n}{L}$ at different $L$'s which merge with each other for $T<T_c$, the merging point is slightly below $T_c$, thus determining the merging point in the thermodynamic limit is not trivial. Instead we choose a value $R$ smaller but not too far from the critical value of $\frac{\xi_n}{L}$ such that $\frac{\xi_n(T)}{L}=R$ has a solution $T_c(L)$ which is larger than $T_c$. To extrapolate $T_c$ in the limit $L\to\infty$, we conduct the nonlinear fitting \cite{Surungan2005}
\begin{equation}
 T_c(L) = T_c + \dfrac{c^2 T_c}{(\ln bL)^2}.
\end{equation}
To choose the optimized value of $R$, we examine the dependence of $T_c$ on $R$. Typically $T_c$ is weakly dependent on $R$ and there is $R_c$ where $T_c$ reaches maximum. Thus we choose several values of $R$ around $R_c$ to improve the extrapolation.

\bibliography{references}

\end{document}